\input{aipcheck}

\documentclass[
    ,final            
  ]
  {aipproc}

\layoutstyle{6x9}

\begin{document}

\title{Slow Galaxy Growth within Rapidly Growing Dark Matter Halos}


\author{Michael J. I. Brown}{
  address={School of Physics, Monash University, Clayton, Victoria 3800, Australia}
}

\author{the Bo\"otes field collaborations}{}


\begin{abstract}
In cold dark matter cosmologies, the most massive dark matter halos undergo rapid growth between a redshift of $z=1$ 
and $z=0$, corresponding to the past 7 billion years of cosmic time. There is thus an expectation that the 
stellar masses of the most massive galaxies will also rapidly grow via merging over this redshift range. 
While there are examples of massive merging galaxies at low redshift, recent observations 
show that the stellar masses of massive galaxies have grown by only $\sim 30\%$ since $z=1$. 
Some of the literature claims that the slow growth of massive galaxies is contrary to the $\Lambda$CDM paradigm,
although this is not necessarily the case. To determine why massive galaxies are not growing rapidly,
we have modeled how galaxies populate dark matter halos. To do this, we have 
measured the space density and spatial clustering of redshift $z<1$ galaxies in the Bo\"otes field of the 
NOAO Deep Wide-Field Survey. We have then modeled the observations using the halo occupation distribution (HOD) 
formalism. We find that the stellar masses of the largest galaxies are proportional to dark matter halo mass 
to the power of a third. In the most massive dark matter halos, we also find that the stellar mass is distributed 
mostly among "satellite" galaxies. As a consequence, the stellar masses of large galaxies are expected to increase  
relatively slowly, even though they reside within rapidly growing dark matter halos.
\end{abstract}

\maketitle


\section{Introduction}

For plausible cold dark matter cosmologies, the most massive (gravitationally bound) halos of dark matter
form when the Universe is relatively young but acquire much of their mass via hierarchical mergers with 
other halos between redshifts of $z=1$ and $z=0$ \citep{pre74,she99,jen01}, corresponding to the past seven billion years. 
As a consequence, there is the expectation that the stellar masses of the most massive galaxies 
will also undergo rapid growth via merging over the same redshift range, and that observations
contrary to this signal a fundamental flaw with the current $\Lambda$CDM paradigm. This expectation
has been reinforced by many of the galaxy formation simulations and models produced since the early 1990s \citep{kau96,del06},
which have predicted rapid growth of massive galaxy stellar masses at redshifts of $z<1$.

\section{Discord}

There is significant discord within the observational literature about how massive galaxies grow
over cosmic time. Some have claimed that massive galaxies have grown by a factor of $\sim 2$ in 
stellar mass since $z=1$, in agreement with the predictions of the many models  \citep{kau96,van05}. 
Others have claimed that there has been no significant growth of massive galaxy stellar masses since $z=1$ 
 \citep{sca07,col09},
and that massive galaxies may undergo pure luminosity evolution (without stellar mass growth) since
their formation. This would be contrary to the $\Lambda$CDM paradigm. 
It is interesting to note that (until recently) relatively few papers have claimed that massive galaxies grow
slowly at $z<1$, rather than rapidly or not at all \citep{mas06,bro07}.

The discord within the astronomical community is not  surprising for a number of reasons. 
For a start, only in the last decade has it been possible to obtain large samples of $z=1$ galaxies
and measure their properties (redshift, luminosity, stellar mass). It is thus not surprising that the 
first generation of studies of massive galaxy growth had limitations that are being identified or
mitigated by later works. Many measurements of ``observable'' properties of galaxies, such as rest-frame
luminosities and colors, are actually dependent upon models including (at a minimum) galaxy 
spectral energy distributions and the spatial distribution of stars within galaxies. Errors in these models
can lead to surprisingly large errors in the observed evolution of galaxies. 

Caveats in galaxy formation models have not been clearly communicated and understood. 
Disagreement between a particular galaxy formation model using a $\Lambda$CDM cosmology and observations
is often interpreted as a flaw with the $\Lambda$CDM itself. 
However, a galaxy formation model must include a number of recipes for describing
the behavior of baryonic matter, including the cooling and gravitational collapse of gas, the formation
of stars, and the heating of gas by astrophysical sources (e.g., supernovae, quasars).
Given the complexity of modeling the baryonic matter, a disagreement between observations 
and a particular model may have little to do with the validity of the $\Lambda$CDM paradigm.
Conversely, agreement between observations and a particular model of $\Lambda$CDM galaxy formation 
does not necessarily validate the $\Lambda$CDM paradigm.

There is strong evidence for massive galaxies being both ancient and continuing to grow via mergers.
This may explain the strong (often bimodal) opinions held by astronomers on the topic of galaxy growth. 
The stellar populations of the most massive galaxies are approximately ten billion years old 
and their current star formation rates are very low \citep{tin68}. However, there are many 
irrefutable examples of massive galaxies at low redshift undergoing mergers \citep{van05}. 
Clearly massive galaxies are undergoing some growth via mergers with other galaxies, 
contrary to pure luminosity evolution models. Thus the debate about galaxy growth should not
be about whether or not there is any growth via mergers. The debate should be about the 
rate of growth via merging.

While one can identify examples of merging galaxies \citep{van05}, it is very hard to turn 
catalogs of merging galaxies into a robust measure of galaxy growth. While some 
merging galaxies can be easily identified, how long they remain recognizable as merging
galaxies is unclear, and this has a direct impact on the measured merger rate.
Even merger rates defined using galaxy pairs depend upon model assumptions for
how long it takes galaxy pairs to merge. Consequently, estimates of 
galaxy growth derived from catalogs of merging galaxies have considerable scatter.

\section{The Slow Growth of Massive Galaxies}

The evolving number of galaxies per unit (comoving) volume is a conceptually 
simple and robust test of how rapidly the stellar masses of massive galaxies 
are increasing. At the redshifts where massive galaxy growth is taking place, the number
of massive galaxies (of fixed stellar mass) per unit volume will increase 
with decreasing redshift.

\begin{figure}
 \includegraphics[height=.40\textheight]{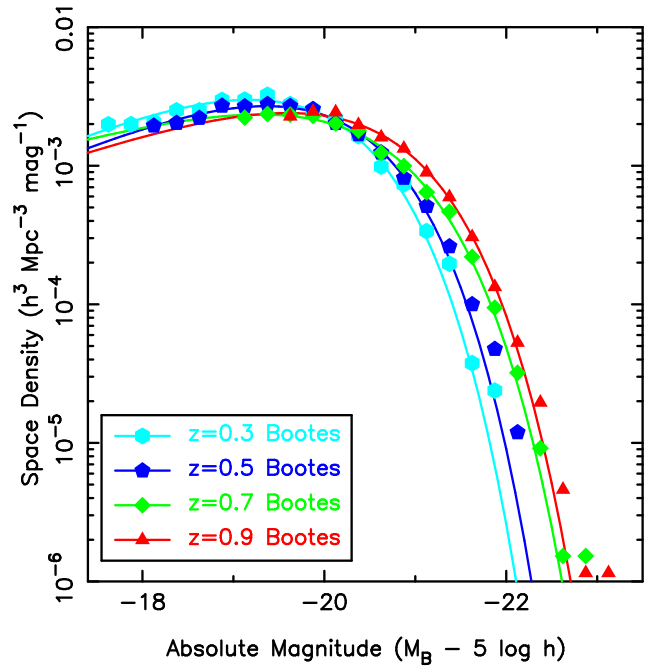}
  \caption{The evolving red galaxy luminosity function \citep{bro07}. The bright end of the luminosity function
(shown at the right of the plot) fades with time, due to the evolution of stellar populations within galaxies. If galaxies were
rapidly growing via merging, we would see the number of very luminous galaxies increasing with time.
The fading of the bright end of the luminosity function does not match stellar population
models, implying some growth of massive galaxies via galaxy mergers. (The parameter $h$ is equal to the
Hubble constant divided by $100~{\rm km~s^{-1}~Mpc^{-1}}$.)}
\label{fig:lf}
\end{figure}

Over the past 5 years, a number of studies measuring the luminosity function
(galaxies per unit volume per unit luminosity) and the stellar mass function (galaxies
per unit volume per unit galaxy stellar mass) have shown that massive galaxies grow
slowly \citep{bun06,bro07,sca07}. 
For example, in Figure~\ref{fig:lf} we plot the evolving red galaxy luminosity function \citep{bro07}.
The bright end of the luminosity function fades with time, largely due to the 
evolution of galaxy stellar populations. If massive galaxies were growing very 
rapidly via merging, this would overwhelm the observed fading, and 
the bright end of the luminosity function would be increasing with time.
While some studies (with large uncertainties) find that the evolution 
of the luminosity function and stellar mass function is consistent with no 
growth of massive galaxies, studies using larger samples of galaxies find
evidence for some galaxy growth. For example, using a sample of $\sim 40,000$ red galaxies, 
we find that the stellar masses
of the most massive galaxies have grown by $30\%$ since $z=1$ \citep{bro07,bro08}. Is the 
slow growth of massive galaxies consistent with $\Lambda$CDM?

\section{Galaxy Growth within Dark Matter Halos}

\begin{figure}
 \includegraphics[height=.3\textheight]{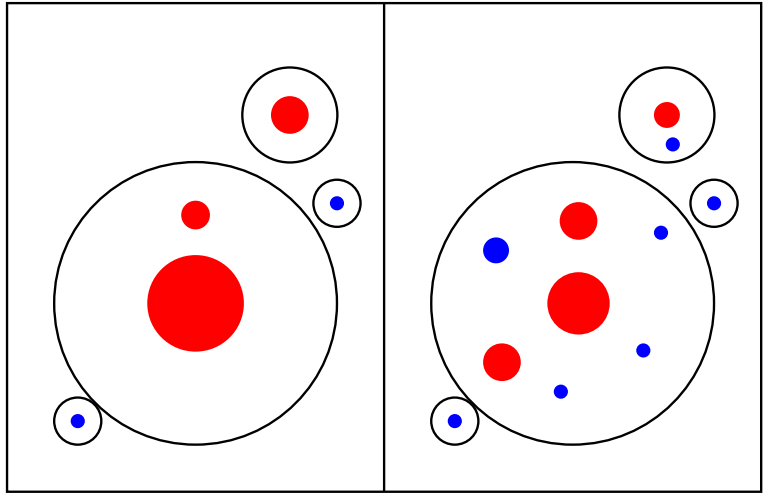}
  \caption{Two schematics illustrating the relationship between galaxy growth and how galaxies
populate dark matter halos. In the first scenario, when dark matter halos (open circles) merge the galaxies (filled circles) 
within them merge soon after. As a consequence, satellite galaxies are relatively rare and galaxy stellar mass is strongly
correlated with dark matter halo mass. In the second scenario, when dark matter halos merge the individual galaxies survive
for many billions of years. As a consequence, satellite galaxies are relatively common and galaxy stellar mass is weakly
correlated with dark matter halo mass.}
\label{fig:scen}
\end{figure}

One can gain insights into why galaxies grow slowly by determining
how galaxies populate extended halos of dark matter. To illustrate this, in Figure~\ref{fig:scen}
we show two scenarios for galaxy growth with the same underlying cosmology. 
In the first scenario, when dark matter halos merge the galaxies that were
within these halos merge soon after. As a consequence, galaxy stellar mass
is strongly correlated with dark matter halo mass and satellite galaxies
are relatively rare. In the second scenario in Figure~\ref{fig:scen}, when dark matter halos 
merge the galaxies orbit within the halo for many billions of years before merging. 
As a consequence, galaxies grow relatively slowly, galaxy stellar mass is weakly
correlated with dark matter halo mass and satellite galaxies are relatively common.
If we can determine how galaxies populate dark matter halos, we can discriminate
between such scenarios.

Unfortunately, it is difficult to determine reliable dark matter halo masses for 
individual galaxies. Stellar and gas kinematics can generally only measure
the mass within the very inner part of the dark matter halo. However, the number
density and spatial distribution of dark matter halos is a function of their mass
and can be accurately modeled (for a given set of cosmological parameters)~\citep{pre74,she99,jen01}. 
As a consequence, one can determine the typical dark matter halo mass for a
given galaxy population by comparing its observed spatial (or angular) clustering with the 
clustering of dark matter halos. As we illustrate in Figure~\ref{fig:scen}, the spatial clustering 
of galaxies (measured using galaxy pairs) on small scales is sensitive to the number of galaxies per dark matter halo. 
One can thus use measurements of the number and spatial clustering of galaxies
to determine how galaxies populate dark matter halos. 

In two papers we have determined how massive galaxies populate dark matter halos \citep{whi07,bro08}. 
To do this, we used 40,696 red galaxies with photometric redshifts between $z=0.2$
and $z=1.0$ in the Bo\"otes field of the NOAO Deep Wide-Field Survey \citep{jan99,eis04}. 
We have determined how red galaxies populate dark matter halos, by utilizing the 
halo occupation distribution (HOD) formalism \citep{pea00,zhe04} to model 
the observed space density and angular clustering of red galaxies.

\begin{figure}
 \includegraphics[height=.35\textheight]{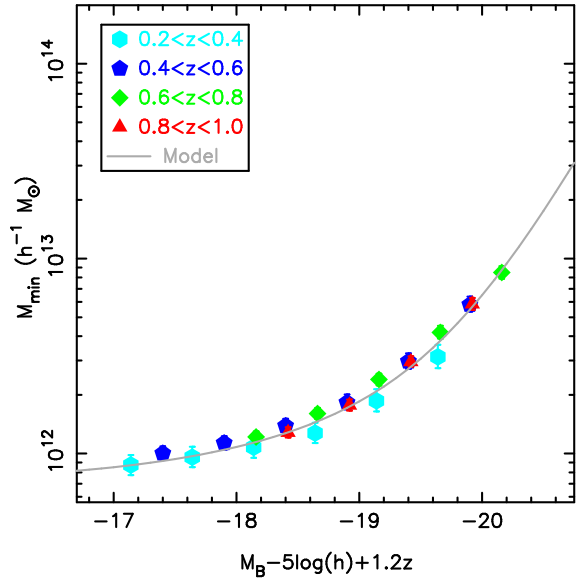}
  \caption{The dark matter halo masses of galaxies as a function of ``evolution corrected'' absolute magnitude \citep{bro08}, 
which is effectively a proxy for stellar mass. Although galaxy stellar masses and host halo masses evolve, the
relationship between these two quantities shows little evolution at $z<1$. In the most massive galaxies, stellar 
mass is proportional to halo mass to the power of a third.}
\label{fig:dm}
\end{figure}

In Figure~\ref{fig:dm} we show the relationship between dark matter halo 
mass and a proxy  for central galaxy stellar mass (an evolution corrected galaxy luminosity). 
A remarkable feature of this plot is the relationship between galaxy stellar mass
and dark matter halo mass does not appear to evolve with redshift. There also 
seems to be a lower limit for the dark matter halo masses of red galaxies, 
and lower mass halos presumably host blue galaxies. For the most luminous galaxies,
galaxy stellar mass is proportional to halo mass to the power of a third. This is similar
to the second scenario noted in Figure~\ref{fig:scen}, where galaxies grow slowly within rapidly
growing dark matter halos.

\begin{figure}
 \includegraphics[height=.35\textheight]{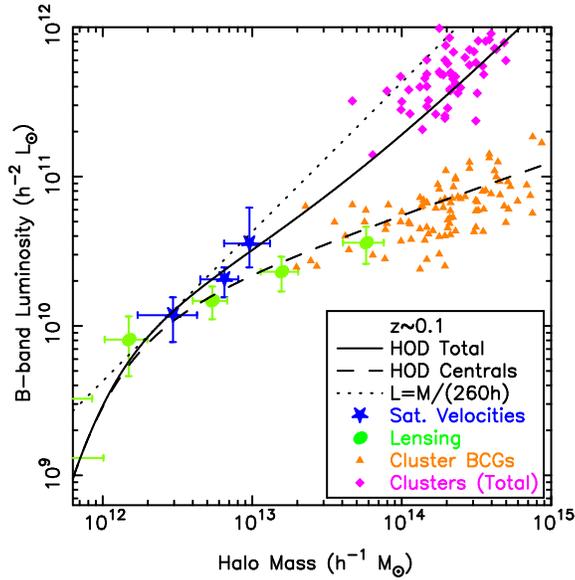}
  \caption{The light produced by central galaxies (dashed line) and the sum of all galaxies (solid line)
as a function of dark matter halo mass \citep{bro08}.
In the highest mass halos, there is more stellar mass contained
in satellite galaxies than within the single central galaxy. Mergers with very massive
dark matter halos do not always lead to rapid growth of massive galaxy stellar masses.
The data points plotted here are not from our study but from the work of others \citep{lin04a,lin04b,man06a,con07a}, 
who derived halo masses using approaches completely independent of the HOD methodology we have employed 
(e.g., weak gravitational lensing).
}
\label{fig:light}
\end{figure}

In Figure~\ref{fig:light}, we show the amount of light produced by galaxies
as a function of dark matter halo mass. As noted previously, central
galaxy stellar mass increases slowly with dark matter halo mass. If one sums the light from 
all galaxies (central and satellites) within halos, one finds that the bulk of the stellar
mass in the most massive halos resides within satellite galaxies, not a single central galaxy. 
Mergers of dark matter halos do not always lead to (immediate) mergers of galaxies. 
It thus appears that the slow growth of massive galaxies may be consistent with 
a $\Lambda$CDM cosmology. However, we caution that this consistency should not
be confused with a robust test of the $\Lambda$CDM paradigm.

\section{Summary}

In CDM cosmologies, the most massive dark matter halos undergo rapid growth via merging at $z<1$.
As a consequence, there has been the expectation that massive galaxies will also rapidly grow
via merging, and that the observed  growth of massive galaxies is a test of the $\Lambda$CDM paradigm. Recent
observations clearly show that the stellar masses of the most massive galaxies do not grow rapidly
at $z<1$. However, this is not necessarily inconsistent with $\Lambda$CDM.
We have investigated how massive galaxies populate dark matter, using halo occupation distribution
modeling of the observed number and clustering of red galaxies. We find that massive galaxy 
stellar masses are proportional to halo mass to the power of (roughly) a third.
We also find that much of the stellar mass within the most massive halos resides within 
multiple satellite galaxies, rather than a single central galaxy.
As a consequence, we find that massive galaxies grow slowly within rapidly growing dark matter halos.
The slow growth of massive galaxies may thus be consistent with $\Lambda$CDM.

\begin{theacknowledgments}
This work is based in part on observations made with the {\it Spitzer} Space Telescope, which is operated
by the Jet Propulsion Laboratory, California Institute of Technology
under a contract with NASA. This research was supported by the National Optical Astronomy Observatory which is
operated by the Association of Universities for Research in Astronomy (AURA), Inc.
under a cooperative agreement with the National Science Foundation.
Several of the key ideas presented here were developed during summer workshops of the Aspen Center for Physics,
who we thank for their hospitality.
\end{theacknowledgments}





\bibliographystyle{aipproc}   

\bibliography{ms}

\IfFileExists{\jobname.bbl}{}
 {\typeout{}
  \typeout{******************************************}
  \typeout{** Please run "bibtex \jobname" to optain}
  \typeout{** the bibliography and then re-run LaTeX}
  \typeout{** twice to fix the references!}
  \typeout{******************************************}
  \typeout{}
 }

\end{document}